# Unconventional Relaxation Dynamics in $Co_8Zn_7Mn_5$ and $Co_8Zn_8Mn_4$: Evidence of Inertial Effects

P. Saha[1], M. Singh[1], P. D. Babu[2] and S. Patnaik[1*]

[1]*School of Physical Sciences, Jawaharlal Nehru University, New Delhi 110067, India*

[2]*UGC-DAE Consortium for Scientific Research Mumbai Centre, BARC Campus, Mumbai 400085, India*

**Corresponding author: spatnaik@jnu.ac.in*

## Abstract

Magnetization relaxation dynamics serve as an essential tool for uncovering the intrinsic mechanisms governing the magnetic response and energy dissipation in magnetic systems. In this work, we examine the relaxation dynamics for β-Mn-type $Co_8Zn_7Mn_5$ and $Co_8Zn_8Mn_4$ across a frequency range of 1 kHz to 10 kHz, spanning different magnetic phases. While most magnetic systems tend to follow the Debye-like relaxation with non-zero distribution or the Cole-Cole formalism, our analysis reveal that these conventional models fail to capture frequency dependence of ac susceptibility across different magnetic phases in $Co_8Zn_7Mn_5$ and $Co_8Zn_8Mn_4$. Instead, an inertial component is needed to successfully describe the dynamics, suggesting the presence of unconventional relaxation behaviour. The characteristic relaxation time (τ) is found to be of the order of $10^{-5}$ s for both the compositions. The field dependent variation of τ exhibits a non-monotonic nature, with the double peak like structure at the skyrmion phase transitions, implying slower relaxation dynamics at the phase boundaries. Furthermore, the presence of non-zero difference between isothermal and adiabatic susceptibility in the pure phases implies slower relaxation dynamics, which is consistent with the presence of finite dissipation in pure phases. The inertial term has been previously invoked to describe the dynamics in spin ice systems due to the propagation of magnetic monopoles. However, it's necessity in this system, points to a wider significance in magnetization dynamics that goes beyond the conventional spin ices and skyrmions. By systematically ruling out well-known mechanisms like monopoles, domain walls and skyrmions, we demonstrate that the inertial effect is an intrinsic property of β-Mn-type Co-Zn-Mn compounds that is independent of any particular magnetic phase. All these findings open a significant new avenue for understanding relaxation processes in chiral magnets and necessitates further studies to investigate the underlying causes of inertial effects in ac susceptibility.

## Introduction

Noncentrosymmetric chiral magnets have attracted research interests in recent decades due to their ability to host non-trivial magnetic textures, such as skyrmions [1]. Magnetic skyrmions are nanometer-sized, vortex-like twisted spin textures that exhibit robust topological protection characterized by a topological winding number. These non-trivial spin textures can be stabilized due to the presence of different mechanisms such as Dzyaloshinskii Moriya (DM) interaction [2], magnetic dipolar interaction [3], frustrated exchange interaction [4] and four spin exchange interactions [5]. The scientific interests in these topological spin configurations stems from their exhibition of unique electromagnetic phenomena such as skyrmion magnetic resonance [6], topological Hall effect [7] and thermally controlled ratchet motion [8]. Notably, skyrmion motion can be induced by a threshold current density of roughly $10^6$ A/m$^2$, which is 5–6 orders of magnitude lower than what is needed to move magnetic domain walls in ferromagnets [9,10]. This behaviour has been attributed to the topological feature of skyrmions [10]. These properties make skyrmions promising candidates for potential usage in various spintronic devices and as low current, high density information carriers [11-13].

Of the different types of skyrmions, the ones stabilised by the competition between the Heisenberg exchange and the Dzyaloshinskii Moriya interaction (DMI) are particularly intriguing because of the small size (<150 nm), making them suitable for high density storage applications [14]. These type of skyrmions have been experimentally observed in non-centrosymmetric B20 compounds [2,15-17] and are stabilized in the form of elongated tubes in a specific field and temperature window just below the Curie temperature ($T_c$). However, the extremely low Curie temperature of B20 compounds make them unsuitable for practical applications. A new family of DMI-induced skyrmions hosting chiral magnets known as β-Mn-type $Co_xZn_yMn_z$ (x+y+z=20) compounds have emerged as the most promising candidates to address this issue due to their large Curie temperature $T_c$, which ranges from 148 to 462 K, depending on the Mn composition [14, 18]. Recent discovery of robust metastable skyrmions in $Co_8Zn_8Mn_4$ over a broad range of temperature and magnetic field have offered a promising platform for technological applications [19]. Electrical manipulation of skyrmions have been realized in $Co_8Zn_{10}Mn_2$ and $Co_9Zn_9Mn_2$ through current pulses [20, 21]. This has paved way to skyrmion applications in memory or logic devices [20, 21]. As a result, $Co_xZn_yMn_z$ compounds have become a primary focus of research in the past decade [19, 22-23].

Although significant research has been done on the $Co_xZn_yMn_z$ compounds, but most part of the studies focus on examining the skyrmion phase using sophisticated methods such as Lorentz Transmission Electron Microscopy (LTEM) [25] and Small Angle Neutron Scattering [26]. Nevertheless, ac susceptibility offers an alternative approach since it is a promising tool for identifying different magnetic phases and studying the relaxation dynamics of magnetization in different systems. By analyzing the ac susceptibility behaviour at different frequencies and applied ac fields, it is possible to probe different magnetic ordering such as ferromagnetic, ferrimagnetic, antiferromagnetic, spin glass phase and super-paramagnetism [27, 28]. In many recent studies, frequency dependent ac susceptibility has been used to study the relaxation dynamics of magnetic modulated phases in skyrmion hosting compounds [29-32]. These studies are important not only to obtain the variation in relaxation frequency across different magnetic phases but also for determining the presence of intricate underlying mechanisms [30,31].

In most of the skyrmion hosting systems, the Cole-Cole model which is an extension of the Debye relaxation model has been employed to describe the frequency dependent behaviour of the ac susceptibility [29-32]. However, in some cases, the frequency dependent behaviour of either the real or the imaginary part or both have shown deviation from the conventional Cole-Cole model [29-31]. For example, in $Mn_{1.4}PtSn$, the out-of-phase component of ac susceptibility (χ'') showed a deviation from the Debye-like relaxation in the high frequency region which was attributed to the eddy current losses [29]. Similarly, in $Fe_{1-x}Co_xSi$, the Cole-Cole formalism involving a distribution of relaxation times has failed to adequately fit the frequency dependent behaviour of in-phase component of ac susceptibility (χ') and χ'' [30]. This discrepancy has been attributed to the presence of several co-existing relaxation processes [30]. Such departure from conventional models have necessitated the need for alternative approach, to explain the dynamics precisely.

Here, we present a systematic investigation of frequency dependent ac susceptibility of β-Mn-type $Co_xZn_yMn_z$ (x+y+z=20) compounds, a class of materials different from the B20 helimagnets in terms of crystal structure, magnetic interactions and temperature range for the stabilization of skyrmion phase [20]. Owing to their high Curie temperature, the relaxation dynamics has been probed using a frequency range between 1kHz to 10kHz, focussing on two compositions: $Co_8Zn_7Mn_5$ and $Co_8Zn_8Mn_4$, the former being an unexplored composition with no earlier reports. Our study reveals that the frequency dependent ac susceptibility study in $Co_8Zn_7Mn_5$ and $Co_8Zn_8Mn_4$ cannot be explained by the conventional Debye relaxation or the Cole-Cole models. Rather, a proper description requires the inclusion of inertial effect in the model. Similar approach has been used previously to explain the dynamics of spin ice systems, where the inertial effects are induced by the magnetic monopoles [33,34].

In the case of this system, the inertial term was applied across the entire field range, covering all magnetic phases from helical to conical to skyrmionic and ferromagnetic. Additionally, incorporating the inertial term not only improved the fit but produced a good agreement between the relaxation time derived from the $\chi$' (f) and $\chi$''(f). The analysis further revealed that the relaxation time is of the order of $10^{-5}$ s across all different magnetic phases. Moreover, field dependent behaviour of relaxation time and the difference between the isothermal and adiabatic susceptibility showed a non-monotonic nature, with maximas occurring near the skyrmion phase transition boundaries.

These findings pose core questions about the genesis of inertial effects in ß-Mn type Co-Zn-Mn compounds implying that the inertial contribution is an intrinsic property instead of a phase-dependent behaviour. This challenges the conventional models of magnetic relaxation and underscores the requirement for a deeper understanding of relaxation dynamics beyond the spin ice systems.

## A. Experimental Techniques

The $Co_8Zn_8Mn_4$ sample used in this work is the same as the one reported in [34]. A conventional melting procedure was used to synthesize the polycrystalline sample of $Co_8Zn_7Mn_5$ using stochiometric amounts of Cobalt, Zinc and Manganese. The mixture was ground for half an hour, after which it was pelletized and vacuum-sealed in a quartz tube. The tube was then heated at 1273K for 15h, followed by slow cooling to 973K at a rate of 1K/h and subsequently quenched in water. The final product was in the form of an ingot with large single crystalline grains. The sample pieces used for measurement had sufficiently large grains which could be manipulated as single crystals. The identification of crystal structure and phase purity was done using a Rigaku X-Ray diffractometer (Miniflex-600). The elemental composition of the synthesized sample was determined using a Bruker AXS microanalyzer. The dc magnetization measurements were carried out using a vibrating sample magnetometer (VSM) probe inserted in a Cryogenic Physical Properties Measurement System (PPMS). A Quantum Design Physical Properties Measurement System (PPMS) was used to perform the ac susceptibility measurements. For the frequency dependent ac susceptibility χ (f) measurement, the sample was initially zero field cooled from above $T_c$ to the desired temperature. Once, the desired temperature was reached, the measurements were performed by sweeping the dc field and the ac field was kept constant at $H_{ac}$=10 Oe. The measurements were taken for a frequency range between 1747Hz to 9747Hz.

## B. Results and Discussions

## Structural Characterization

Main panel of figure 1 displays the powder X-ray diffraction (PXRD) pattern (black curve) of $Co_8Zn_7Mn_5$. The red curve represents the Rietveld refined fit performed using the FullProf software. All the peaks align well with the spacegroup $P4_132$ (no. 213). The lattice parameters yielded from the fit are a = b = c = 6.38 Å, which are comparable in magnitude with other ß-Mn type $Co_xZn_yMn_z$ compounds [18]. Left side of figure 1 shows the cubic crystal structure of $Co_xZn_yMn_z$ compounds generated through VESTA software. As depicted, the unit cell comprises of 20 atoms which are positioned at two crystallographic sites : eightfold site (8c) and twelvefold site (12d). Previous reports have demonstrated that the 8c sites are strongly preferred by the Co atoms while, the 12d sites are randomly occupied by the Zn and Mn atoms [18]. Inset at the right shows the energy dispersive X-ray (EDX) spectrum of the synthesized sample. The chemical composition of the prepared sample is $Co_{7.94}Zn_7Mn_5$, which is very similar to the nominal stoichiometry.

## Dc Suceptibility

Figure 2 shows the variation in dc susceptibility with temperature at 100 Oe in $Co_8Zn_7Mn_5$. The measurements were taken in two different protocols: zero field cooled warming (ZFW) and field cooled warming (FCW). Figure 2 shows a sharp rise in susceptibility in both the protocols (ZFW and FCW) near the Curie temperature ($T_c$), which marks the onset of the magnetic phase transition [36,37]. Curie temperature ($T_c$) can be obtained from the minima of the $d\chi/dt$ vs T behaviour. Inset of figure 2 shows the $T_c$ for $Co_8Zn_7Mn_5$ to be at 252K. Similar to the other compositions of $Co_xZn_yMn_z$ compounds [18,35,37-39], a wide bifurcation arises in FCW and ZFW protocols. The divergence between these two curves originate from the varying magnetization between the aligned and the misaligned domains in case of FCW and ZFW. As seen in figure 2 dc susceptibility tends to drop gradually below 125K. This downturn in dc susceptibility is more pronounced in Mn rich compositions of Co-Zn-Mn compounds [18, 38, 42] and can be attributed to the disorder induced by the Mn moments [18] or by the increment of the helical wave vector q [43]. Furthermore, the ZFW curve depict a sharp drop in susceptibility below 25K. This signifies the onset of a fully disordered reentrant spin glass phase which arises due to the freezing of the Mn spins [18]. Here, the phrase "reentrant" describes the transition of Mn spins in Co-Zn-Mn compounds from the paramagnetic phase at high temperature to a partially ordered state at $T_c$ and then into a disordered spin glass state at low temperatures [18]. The inverse dc susceptibility as a function of temperature was analyzed using the Curie–Weiss law, expressed as $\chi = C/(T - \Theta_{cw})$, where C denotes the Curie constant and $\Theta_{cw}$ represents the Curie–Weiss temperature [40]. A positive $\Theta_{cw}$ value of approximately 249 K indicates that $Co_8Zn_7Mn_5$ exhibits ferromagnetic behaviour.

## Skyrmion phase

Leveraging the sensitivity of ac susceptibility to various modulated magnetic phases, χ' was measured as a function of the dc magnetic field at different temperatures at a constant frequency of 347Hz. Figure 3(a) represents the field dependent ac susceptibility χ' for a temperature range of 236K- 250K for $Co_8Zn_7Mn_5$. The peaks and dips associated with the χ' behaviour corresponds to different magnetic phases as described in [37]. Specifically, the χ' initially rises with increasing field and this is associated with the change in magnetic phase from the helical to conical. Subsequently, a small dip in the χ' is observed for a temperature range of 239K-246K for $Co_8Zn_7Mn_5$. This pocket like anomaly existing between two peaks is a common characteristic of skyrmion hosting compounds [30-32, 37-38]. In fact, this double-peak structure corresponds to the transition from conical (high susceptibility) to skyrmion phase (low susceptibility) and back to the conical phase (high susceptibility). For temperatures below 238K in $Co_8Zn_7Mn_5$ the dip corresponding to the skyrmion phase is absent; however, the helical to conical transition indicated by the increase in χ' is still observed. To determine the precise phase boundaries of the compound, which could not be identified through smooth phase transitions, the derivative of χ' with respect to the magnetic field (dχ'/dH) was calculated. The inflection points, corresponding to the local maxima and minima, were then used to pinpoint the phase transition boundaries as in [45]. Figure 3(b) illustrates this approach at 241 K where all the different magnetic phases (helical, conical and skyrmion) are present. Plotting these points on the temperature and field axes yields a magnetic phase diagram (figure 3(c)) that is in agreement with the phase diagram of other compositions of $Co_xZn_yMn_z$ compounds. This commonly observed phase diagram shows a helical ground state, that transitions into a canted conical state with increasing field and finally transforms into a field polarized ferromagnetic state. Here, skyrmion phase temperature window is 10K wide in $Co_8Zn_7Mn_5$. This is the first report on the magnetic phase diagram of $Co_8Zn_7Mn_5$.

## Frequency dependence of χ' and χ"

Figure 4 shows the variation in the in-phase and out-of-phase components of ac susceptibility (χ' and χ") with changing dc field over a frequency range of 1747Hz to 9747Hz. These measurements were taken after zero field cooling from above $T_c$ to a constant temperature: 240K for $Co_8Zn_7Mn_5$ and at 265K for $Co_8Zn_8Mn_4$ where all the distinct magnetic phases are present. The dip like anomaly which is the characteristic of the skyrmion phase is present in both field dependent χ' and χ" of $Co_8Zn_7Mn_5$ and $Co_8Zn_8Mn_4$ as shown in Figure 4. To comprehensively investigate the change in susceptibility with frequency for the whole range of magnetic fields covering the helical, conical, skyrmionic, and ferromagnetic states, the χ'(H) and χ"(H) have been replotted as a function of frequency as shown in Figure 5. Figure 5(a) and 5(c) show that χ' varies significantly with frequency for all the field ranges, clearly indicating that the relaxation frequency in chiral magnet $Co_8Zn_7Mn_5$ and $Co_8Zn_8Mn_4$ lies close the experimental frequency window. Notably, the requirement for higher frequencies to probe the relaxation dynamics in this system stands in stark contrast to other skyrmion-hosting compounds such as $Fe_{1-x}Co_xSi$, $GaV_4S_8$, and $Cu_2OSeO_3$, where a lower frequency range of 0.1–1 kHz is typically sufficient [30-32]. The requirement for higher frequencies to probe the relaxation dynamics in $Co_8Zn_7Mn_5$ can be attributed to its high Curie temperature ($T_c$), which is higher than most of the previously studied skyrmion-hosting compounds [30-32]. This observation is consistent with the behaviour of $Mn_{1.4}PtSn$, where a high $T_c$ (~400K) corresponds to a large exchange energy (J), leading to shorter switching times as $\tau$ is proportional to $\hbar/J$ [29]. Consequently, in systems with large J, like $Co_8Zn_7Mn_5$, higher frequencies are needed to investigate the relaxation dynamics.

Figure 5(b) and (d) show a growing trend of χ" with frequency, but the rate of growth eventually decreases at higher frequencies, indicating a potential saturation. In general, the χ" exhibits a peak as the experimental frequency approaches the characteristic relaxation frequency of the system. The observed trend of increasing χ" at a slower rate suggests that the characteristic relaxation frequency of the material is beyond the high-frequency range of the experimental window. As the applied ac field changes as a function of time, the moments in the sample try to align itself with the applied field. But due to the presence of competing physical mechanisms in the system, internal friction can occur which can lead to dissipation or loss of energy. χ" reflects this energy dissipation in a system. In most skyrmion-hosting compounds, χ" remains zero across different magnetic phases, except at skyrmion phase transition boundaries (conical to skyrmionic and skyrmionic to conical) [30-32]. However, in $Co_8Zn_7Mn_5$ and $Co_8Zn_8Mn_4$, the χ" is non-zero for all different magnetic phases with maximas occurring at the skyrmion phase transitions. This nonzero behavior implies that energy dissipation in these two compounds is not just restricted to skyrmion phase boundaries, but occurs across all magnetic phases. The fluctuation-dissipation theorem provides a direct relationship between a system's inherent fluctuations and its energy dissipation [45]. It has already been suggested that the thermal fluctuations play a crucial role in stabilization of the skyrmion phase near $T_c$ [46]. Furthermore, the muon spin relaxation experiments on Co-Zn-Mn family of compounds have revealed a broad peak in muon spin relaxation rate λ even at fields and temperatures outside the skyrmion phase window [47]. The large λ reveals the presence of strong magnetic fluctuations in this system [45]. As a result, the consistent high value of χ" through all magnetic phases implies the persistent influence of magnetic fluctuations in the system. Moreover, as the system approaches the ferromagnetic phase, $\chi''$ decreases monotonically. This is due to the suppression of magnetic fluctuations with increasing field, resulting in less energy dissipation.

In order to study the relaxation dynamics of a magnetic system, Casimir and du Pré [48] derived a model for ac susceptibility which was analogous to the Debye model of dielectric relaxation. According to this model, the complex ac susceptibility is expressed as follows:

$$\chi(\omega) = \chi_s + \frac{\chi_0 - \chi_{inf}}{1 + i\omega\tau_o} \quad (1)$$

where $\chi_0$ and $\chi_{inf}$ are the isothermal and adiabatic susceptibilities that can be expressed as $\chi(0)$ and $\chi(\infty)$. $\omega$ is the angular frequency $(2\pi f)$ and $\tau_o$ is the characteristic relaxation time $(1/2\pi f_o)$ of the system under study. The above equation is applicable for systems where the magnetization relaxes exponentially with a single relaxation time [28]. In order to account for the systems exhibiting a spread in the relaxation times, a parameter α was introduced leading to the formation of a new model known as the generalized Debye model or the Cole-Cole relation [49]:

$$\chi(\omega) = \chi_s + \frac{\chi_0 - \chi_{inf}}{1 + (i\omega\tau_o)^{1-\alpha}} \quad (2)$$

Where α corresponds to the measure of distribution of relaxation time with the value of α ranges between zero (no distribution) to 1 (infinite distribution). Nevertheless, neither model was able to adequately represent the observed frequency-dependent behavior $\chi'$ and $\chi''$ in $Co_8Zn_7Mn_5$ and $Co_8Zn_8Mn_4$. Additionally, the relaxation times derived from fitting the $\chi'$ and $\chi''$ were an order of magnitude different, indicating the need for a more refined approach to adequately depict the behavior of these systems. This led to the exploration of the alternative approaches to better explain the relaxation dynamics in $Co_8Zn_7Mn_5$ and $Co_8Zn_8Mn_4$.

To address these discrepancies, an alternative model inspired by the work of N.P Armitage et al was considered [33]. Here, they introduced a new relaxation model for incorporating the presence of inertial effect in the system by adding a term corresponding to the second time derivative of the coordinate of interest [33]. This method provides a more thorough description of the relaxation process by taking into consideration the impact of inertia on the dynamics of the system. Despite the possibility of different physical origins, incorporating the inertial term produced a significantly better fit, suggesting its significance in describing the relaxation behavior of $Co_8Zn_7Mn_5$ and $Co_8Zn_8Mn_4$. To represent this behaviour, the following modified relaxation equation has been employed [33]:

$$\chi(\omega) = \chi_{inf} + \frac{\chi_0 - \chi_{inf}}{1 + (i\omega\tau_o)^{1-\alpha} - \frac{\omega^2}{\omega_o^2}} \qquad (3)$$

Here, the above equation differs from the Debye-like relaxation (equation 2) as here the response function falls faster with frequency due to the presence of the inertial term $\frac{\omega^2}{\omega_o^2}$. In this case, $\omega_o$ is the characteristic frequency which is related to the inertial reaction of the system. The real and imaginary part extracted from equation 3 are written as follows:

$$\chi'(\omega) = A \frac{1 + (\omega\tau_o)^{1-\alpha}\cos\left(\frac{\pi}{2}(1-\alpha)\right) - \frac{\omega^2}{\omega_o^2}}{[1 + (\omega\tau_o)^{1-\alpha}\cos\left(\frac{\pi}{2}(1-\alpha)\right) - \frac{\omega^2}{\omega_o^2}]^2 + [(\omega\tau_o)^{1-\alpha}\sin\left(\frac{\pi}{2}(1-\alpha)\right)]^2} + \chi_{inf} \qquad (4)$$

$$\chi''(\omega) = A \frac{(\omega\tau_o)^{1-\alpha}\sin\left(\frac{\pi}{2}(1-\alpha)\right)}{[1 + (\omega\tau_o)^{1-\alpha}\cos\left(\frac{\pi}{2}(1-\alpha)\right) - \frac{\omega^2}{\omega_o^2}]^2 + [(\omega\tau_o)^{1-\alpha}\sin\left(\frac{\pi}{2}(1-\alpha)\right)]^2} \qquad (5)$$

Where $A = \chi_T - \chi_{inf}$, represents the difference between the isothermal and adiabatic susceptibilities.

First, we study the frequency dependent behaviour of $\chi'$ and $\chi''$ in $Co_8Zn_7Mn_5$ using equations 3 and 4 respectively. Figure 6(a) and 6(b) depict the corresponding fits to the frequency dependent behaviour of $\chi'$ and $\chi''$at different constant fields at 240 K. Equation 4 accurately describes the behavior of $\chi'(f)$ in $Co_8Zn_7Mn_5$ for all applied fields. Although equation 5 accurately reflects the overall trend of $\chi''(f)$, there is some discrepancy observed in the low-frequency region below 3 kHz. Using these fits, all the parameters such as $A$, $\tau_o$ and $\alpha$ can be obtained for each particular field. The variation of these obtained parameters as a function of the magnetic field is shown in figure 7(a) and 7(b). Figure 7(a) shows the non-monotonic variation of A as a function of applied dc field. In earlier studies on B20 skyrmion-hosting compounds, A vanished in pure magnetic phases and was nonzero only at the skyrmion phase boundaries [31]. However, in $Co_8Zn_7Mn_5$, A remains nonzero across all magnetic phases (helical, conical, and skyrmionic) and gradually decreases after entering the ferromagnetic phase. The persistent nonzero value of A implies a considerable difference in isothermal and adiabatic ac susceptibility, exhibiting strong magnetic relaxation in the helical, conical, and skyrmion phases. A also indicates the presence of considerable dissipation in the system. This behavior of A validates the occurrence of non-zero $\chi''$ across all pure magnetic phases. Figure 7(a) illustrates the presence of two peaks near the skyrmion phase transition boundaries. This increment in the value of A at these two regions reflect the presence of slower relaxation dynamics. Since, skyrmions are topological entities, their creation and annihilation can be characterized as topological phase transitions. Hence, these phase transitions from conical to skyrmionic or vice versa cannot occur without the formation of topological defects such as monopoles and antimonopoles [32]. The emergence of these defects at the phase transition boundaries can result in the slow dynamics. Furthermore, these phase transition boundaries serve as crossover regions of different magnetic phases due to which irregularities in spin coordination can emerge, substantially slowing down the dynamics [29] and maximizing the difference between $\chi_0$ and $\chi_{inf}$. Similar rise in A at the skyrmion phase boundaries have been observed for other B20 skyrmion hosting compound [31] and in antiskyrmion hosting Heusler compounds [29]. Figure 7(b) shows the irregular variation of $\tau_o$with the applied field. $\tau_o$is found to be of the order of $10^{-5}$ across all the different magnetic phases. Similar to the behaviour of A, $\tau_o$ exhibits a double-peak like structure at the same fields where A exhibits maxima near the skyrmion phase boundaries. The rise in the relaxation time near the skyrmion phase transition boundaries aligns with the formation of topological defects as well as the spin irregularities due to the presence of mixed phases [29,32]. Prior research on the skyrmion hosting compounds have shown that the relaxation time increases sharply at the skyrmion phase transition boundaries across different materials. For instance, in $GaV_4S_8$, $Fe_{0.7}Co_{0.3}Si$ and $Cu_2OSeO_3$, where the helical pitch is relatively short (<50nm), relaxation time exceeding 10 s have been reported. On the other hand, the relaxation time in $Mn_{1.4}PtSn$, which has a significantly larger helical pitch (>100 nm) and hosts antiskyrmions, is approximately $10^{-5}$ s [29]. Similarly, in the case of $Co_8Zn_7Mn_5$, where the helical pitch is more than 100 nm the relaxation time has been determined to be of the order of $10^{-5}$ s. A possible reason for this trend, resides in the nature of skyrmions which extend as tubular structures in bulk materials, with their axis along the direction of the applied magnetic field [49-53]. These tubes merge or terminate to form topological defects such as monopoles or antimonopoles during topological phase transitions [49-53]. Since materials with small helical pitch exhibit high density of skyrmions, the density of the topological defects forming at the skyrmion tube boundaries should also be higher. Given that the rise in the relaxation time is linked to these topological defects, a reduced density of topological defects in longer helical pitch can result in a significantly less slower dynamics. While this pattern suggests a possible link between the helical pitch and relaxation behaviour, further studies are required to directly confirm the significance of topological defect density on relaxation dynamics. Moreover, previous studies have suggested that larger sized skyrmions exhibit smoother magnetic configurations, which results in less damping [54]. Therefore, systems hosting larger skyrmions have far shorter relaxation time at skyrmion phase transition boundaries than smaller ones. The distribution of the relation time $\alpha$ has been obtained to be zero across all the different magnetic phases. This implies that all the entities relax simultaneously regardless of whether they are in helical, conical, skyrmion or ferromagnetic phases. Although the zero distribution of relaxation times might appear unexpected given the existence of two separate magnetic entities (Co and Mn), earlier element-specific x-ray spectroscopy experiments have indicated that Co and Mn pair ferromagnetically and exhibit similar magnetic patterns [55]. This agreement in their magnetic behaviour, may help explain the uniform relaxation behaviour observed across the different magnetic phases. Figure 7 (c) and (d) shows the variation of fitted parameters (A and $\tau_o$) with the applied dc field obtained from fitting the behaviour of $\chi''$(f) using equation 5. Both the parameters show a similar trend, displaying a double peak like structure at the skyrmion phase transition boundaries, which aligns with the presence of slow dynamics at these regions. Moreover, the magnitude of the parameters obtained from the fitting of $\chi''$ (f) is of the same order as those from the fitting of $\chi'$(f). In addition, the distribution of relaxation time has been extracted to be zero, which is consistent with the analysis of $\chi'$(f). These findings further reinforce the validity of this new approach for studying the relaxation dynamics in $Co_8Zn_7Mn_5$.

To substantiate the current study using this modified relaxation model (equation 3), similar analysis was conducted to investigate the relaxation behaviour in $Co_8Zn_8Mn_4$. Equations 4 and 5 have been employed to fit the frequency dependent behaviour of $\chi'$ and $\chi''$ in $Co_8Zn_8Mn_4$ as depicted in figure 8(a) and 8(b). The derived parameters (A and $\tau_o$) have been plotted as a function of applied dc field in figure 9. Both the parameters exhibit two peaks near skyrmion phase transition boundaries as seen in figure 9(a) and (b). These peaks signify the nucleation and destruction of the skyrmion phase through formation of spin irregularities and topological defects and thus leading to slow dynamics similar to the behaviour observed in $Co_8Zn_7Mn_5$. Similarly, no distribution in relaxation time has been observed in $Co_8Zn_8Mn_4$, consistent with the current findings for $Co_8Zn_7Mn_5$. The fitted parameters obtained from the fitting of $\chi'(f)$ and $\chi''(f)$ are similar in magnitude. It is to be noted that all the fitted parameters derived from the analysis of $\chi'(f)$ and $\chi''(f)$ in $Co_8Zn_8Mn_4$ lie within the same range as those obtained for $Co_8Zn_7Mn_5$. This similarity in magnitude in both the compounds reflect the presence of similar relaxation dynamics in these systems.

The fitting of $\chi'$ in both $Co_8Zn_8Mn_4$ and $Co_8Zn_7Mn_5$ yields the parameter $\omega_o$, which stays constant at 0.101 MHz throughout the whole field range, encompassing all magnetic phases. Similarly, the $\omega_o$ derived from the fitting of $\chi''$ is determined to be 0.075MHz for $Co_8Zn_7Mn_5$ and 0.082MHz for $Co_8Zn_8Mn_4$ which is again constant across different magnetic phases. This consistency suggests that rather than being related to a particular phase transition or topological excitation, the existence of an inertial term is an inherent characteristic of the ß-Mn type Co-Zn-Mn system. Our results show that the family of β-Mn type Co-Zn-Mn compounds does not adhere to the conventional Debye-like relaxation. Instead, $\chi'(f)$ shows a substantial decline at higher frequencies unlike in other skyrmion hosting samples where the decrease with frequency is more gradual [30-32]. This behaviour can only be explained by incorporating an inertial term in the Debye relaxation model.

Next, we examine the physical processes that support the necessity of inclusion of an inertial factor while describing the relaxation dynamics of the compounds. The inertial effect in the Debye-like relaxation behaviour has been applied previously in literature to understand the dynamics of quantum spin ices [33,34]. In these systems, defects in the magnetic texture, referred to as "monopoles" exhibit properties similar to fundamental magnetic monopoles. The inertial effects in spin ices are induced due to the propagation of these monopoles which move as massive quasi-particles through the lattice [33,34]. The presence of inertial effects is manifested as the change in the sign of $\chi'$ at high frequencies (THz) [33].

The incorporation of an inertial term to describe the frequency dependent behaviour of ac susceptibility in $Co_8Zn_7Mn_5$ and $Co_8Zn_8Mn_4$ sheds light on the microscopic processes that drive the relaxation dynamics in these systems. Through the fitting of both the real and imaginary components of ac susceptibility, especially in the kHz frequency range, suggests that the system has non-trivial dynamic behaviour. Notably, the necessity for an inertial term in spin ices emerged at the THz frequency scale [33], suggesting the involvement of different underlying mechanisms in the current scenario.

Analogous to the monopoles in quantum spin ice, topological defects can emerge in chiral ferromagnets [50-52]. Skyrmions exist in three-dimensional metallic systems as tubular structures called skyrmion lines, which are associated with a quantized magnetic flux [48-51]. Hence, the creation or destruction of these skyrmions are associated with the proliferation of sources or sinks of emergent magnetic flux [50-53]. These points of singularities are identified as topological defects or "magnetic monopoles" which tend to move across the sample along the skyrmion line [56]. Given the similarity between the magnetic monopoles in spin ice systems and chiral ferromagnets, it is plausible that these topological defects can lead to emergence of inertial effect in $Co_xZn_yMn_z$ compounds. However, since these monopoles are created only during the skyrmion phase transition, whereas the inertial term has been applied across all the magnetic phases and not just at the topological phase transition boundaries, the significance of these monopoles in the observed inertial effects in $Co_xZn_yMn_z$ compounds can be ruled out.

Domain walls, which separate the regions of different orientation of magnetization exhibit certain inertia and damping due to their motion under an applied magnetic field [33,57]. With an applied field, the energy of the domain walls are proportional to the square of their velocity [33]. This mechanism was theoretically proposed by Doring [58] which was later experimentally confirmed by Perekalina et al [59]. Therefore, domain wall inertia can be the source of the inertial effects observed in compounds belonging to the ß-Mn type $Co_xZn_yMn_z$ family. The field-dependent magnetization behavior of $Co_8Zn_7Mn_5$ and $Co_8Zn_8Mn_4$ indicates that these systems do not fully saturate at such low fields (0 to 500 Oe) and hence indicates to the presence of domain walls in this regime. This observation raises the idea that the inertial behavior seen in the susceptibility measurement can be caused by domain wall dynamics. Notably, relaxation dynamics have also been investigated in field regimes below the saturation limit in other skyrmion-hosting systems, where domain walls exist but no inertial behavior has been documented [31]. This implies that although domain wall motion might play a role in the relaxation process in $Co_xZn_yMn_z$, it might not be the main cause of the observed inertial effects. Domain walls could therefore be a significant but not a sufficient explanation.

It has also been proposed that time-dependent magnetic forces and thermal fluctuations can induce inertial effects in skyrmion dynamics which can be described by the large effective mass of the skyrmions in chiral magnets [54]. However, if this were the primary origin of the inertial behaviour in $Co_xZn_yMn_z$ compounds, it would be expected to occur only within the magnetic field range that stabilizes the skyrmion phase. Provided, that the inertial behaviour is observed for the entire field range, beyond the skyrmions phase, this explanation is highly unlikely. Hence, the introduction of inertia to study the relaxation dynamics of $Co_xZn_yMn_z$ compounds cannot be attributed to conventional mechanisms such as monopole dynamics or domain wall motion. Among the skyrmion-hosting compounds, $Co_xZn_yMn_z$ compounds are known to exhibit substantial spin fluctuations, as reported in prior studies [18,35,37,60-61]. For instance, the critical behaviour analysis of $Co_7Zn_8Mn_5$ have suggested the presence of strong spin fluctuations [60]. It has been reported that in $Co_xZn_yMn_z$ family of compounds, the Mn moments continue to fluctuate below the ordering temperature and remain persistent even in the presence of strong magnetic fields [18]. These fluctuating Mn spins gradually disrupt the overall magnetic interactions, including those experienced by the Co atoms [18]. This results in a ferromagnetic state which is infact dynamic and fluctuating exhibiting high entropy [18]. Additionally, the analysis of the non-reciprocal transport phenomena in $Co_8Zn_9Mn_3$ has highlighted the role of spin fluctuations contributing to the emergence of non-reciprocal scattering [61]. Moreover, the recent transport studies done on $Co_9Zn_9Mn_2$ and $Co_8Zn_8Mn_4$ have identified the presence of spin fluctuations below $T_c$ [35,37]. All of these findings clearly demonstrate that spin fluctuations are an inherent property of $Co_xZn_yMn_z$ compounds and play an important role in shaping their magnetic and transport behaviour [18,35,37,60-61].

Previous study have shown theoretically that spin fluctuations can impact the response function of materials such as conductivity and damping [62]. A recent study on $Sr_2RuO_4$ showed that the magnetic response function is dominated by spin fluctuations, suggesting that fluctuating spin configurations can modify the dynamical properties in systems exhibiting strong electron correlations [63]. Furthermore, functional renormalization group investigations have demonstrated that spin fluctuations can influence magnon damping, underscoring the relevance of spin fluctuations in transforming the effective equations of motion for spin dynamics [64]. Similarly, theoretical study on asymmetric spin-orbit interactions has showed that critical spin fluctuations can deform the Fermi surface and can cause mass renormalization of quasiparticles [65]. These reports collectively indicate that spin fluctuations can have a significant impact on response functions in magnetic systems, altering both effective mass and dynamics of the system.

In light of these well-established effects, we suggest that an emergent inertial term in the dynamics of AC susceptibility relaxation could result from significant spin fluctuations in $Co_xZn_yMn_z$ compounds. These fluctuations can modify the equation of motion of the relaxation dynamics, by including an inertial term even in the kHz frequency range. Since these measurements have been conducted below the monodomain state, the domain walls are also likely to contribute to the inertial behaviour. Therefore, we cannot completely rule out a partial contribution from domain wall dynamics, even though spin fluctuations seem to play a dominant role. To validate this idea, further theoretical and experimental investigations are needed. Additionally, a better understanding of how spin fluctuations contribute to inertia in ac susceptibility may be possible by theoretical modelling that takes into account fluctuations-induced modifications to relaxation dynamics.

## Conclusion

In conclusion, the analysis of the magnetic relaxation dynamics of β-Mn-type $Co_8Zn_7Mn_5$ and $Co_8Zn_8Mn_4$ revealed a non-Debye type relaxation that could only be captured through the inclusion of inertial effects. The consistency of the inertial term across different magnetic phases implies that these effects are an inherent feature of this material. The characteristic relaxation time extracted is of the order of $10^{-5}$ s. The change in relaxation time and the difference between the isothermal and adiabatic susceptibility across different phases exhibit a non-monotonic nature, peaking at the skyrmion phase transition, indicating slower relaxation dynamics. The difference between the isothermal and adiabatic susceptibility remains non-zero even within pure phases, indicating strong dissipation, which is further supported by the non-zero $\chi''$ across all phases. This dissipation, even in pure phases, points toward the presence of unconventional effects in the system's dynamics. By excluding conventional mechanisms such as monopoles, domain walls, and skyrmions, we suggest that strong spin fluctuations, inherent to the β-Mn-type crystal structure, could be at the root of the inertial effects observed in ac susceptibility measurements.

**Caption for Figures:**

**Figure 1:** Main panel shows the powder X-Ray diffraction data of $Co_8Zn_7Mn_5$. Red line shows the Reitveld refinement fit. Inset at the left shows the cubic crystal structure of ß-Mn type $Co_xZn_yMn_z$ (x+y+z=20) compounds where the blue coloured spheres represent the 8c sites and the green coloured spheres represent the 12d sites. Inset at the right shows the energy dispersive X-ray spectrum of the synthesized sample.

**Figure 2:** shows the change in dc susceptibility as a function of temperature. The black curve corresponds to the zero field cooled warming (ZFW) and the red curve corresponds to field cooled warming (FCW) at 100 Oe. Inset shows the temperature dependent derivative of dc susceptibility which exhibits a minima at $T_c$ ~ 252K.

**Figure 3:** (a) shows $\chi'$ as a function of applied dc field at different temperatures from 236K to 250K for $Co_8Zn_7Mn_5$ at constant frequency of 347Hz. The dip-like anomaly indicates the presence of the skyrmion phase in this compound. (b) shows the variation in $\chi'$, $\chi''$ and $d\chi'/dH$ with respect to applied dc field at 241K. The straight black lines pinpoint the phase boundaries during transition. (c) represents the magnetic phase diagram of $Co_8Zn_7Mn_5$.

**Figure 4:** shows $\chi'$ and $\chi''$ as a function of applied dc field for different frequencies ranging from 1747Hz to 9747 Hz, measured after zero field cooling to a constant temperature for (a) and (c) at 240K for Co8Zn7Mn5 and (b) and (d) at 265K for Co8Zn8Mn4.

**Figure 5:** $\chi'(H)$ and $\chi''(H)$ have been replotted as a function of frequency ($\chi'(f)$ and $\chi''(f)$) for different constant fields: (a) and (b) for $Co_8Zn_7Mn_5$ and (c) and (d) for $Co_8Zn_8Mn_4$.

**Figure 6:** shows the (a) $\chi'(f)$ and (b) $\chi''(f)$ for $Co_8Zn_7Mn_5$ and the red curve curve represent the corresponding fit using equation (4) and (5).

**Figure 7:** shows the fitting parameters: A and $\tau_o$ as a function of magnetic field for $Co_8Zn_7Mn_5$ obtained by fitting (a) and (b) $\chi'(f)$ and (c) and (d) $\chi''(f)$ using equations (4) and (5).

**Figure 8:** shows the (a) $\chi'(f)$ and (b) $\chi''(f)$ for $Co_8Zn_8Mn_4$ and the red curve curve represent the corresponding fit using equation (4) and (5).

**Figure 9:** shows the fitting parameters: A and $\tau_o$ as a function of magnetic field for $Co_8Zn_8Mn_4$ obtained by fitting (a) and (b) $\chi'(f)$ and (c) and (d) $\chi''(f)$ using equations (4) and (5).

Figure 1:

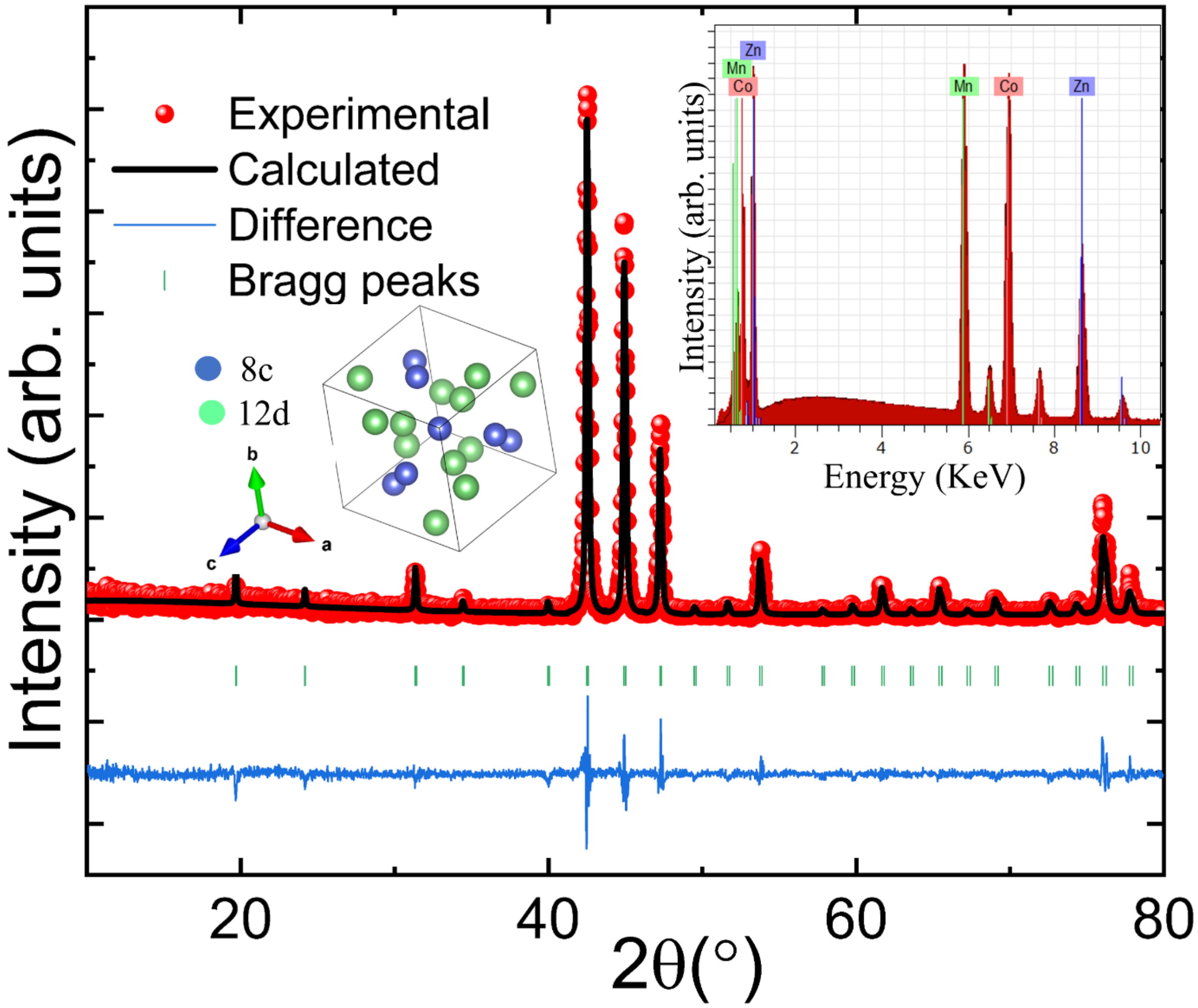

Figure 2:

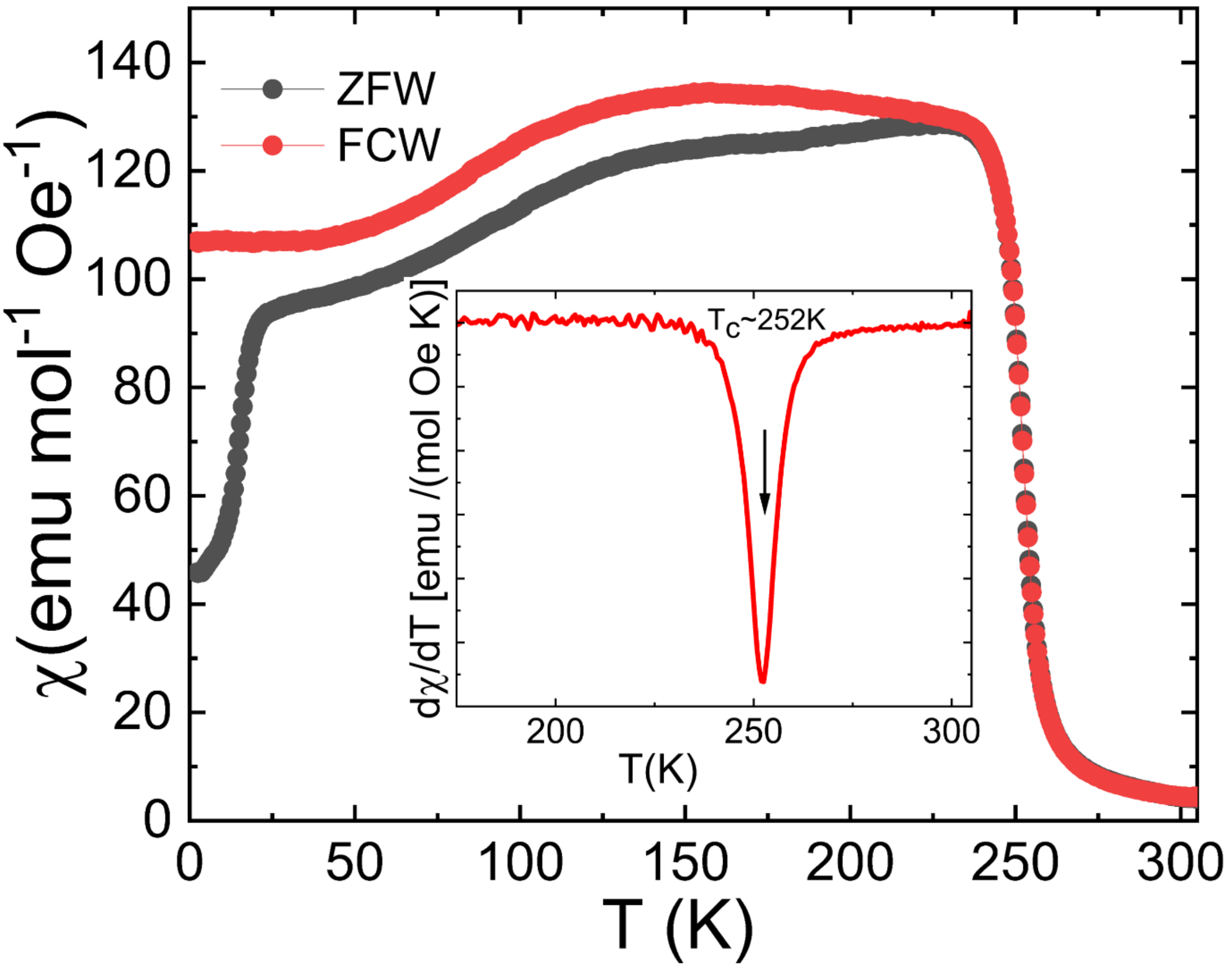

Figure 3:

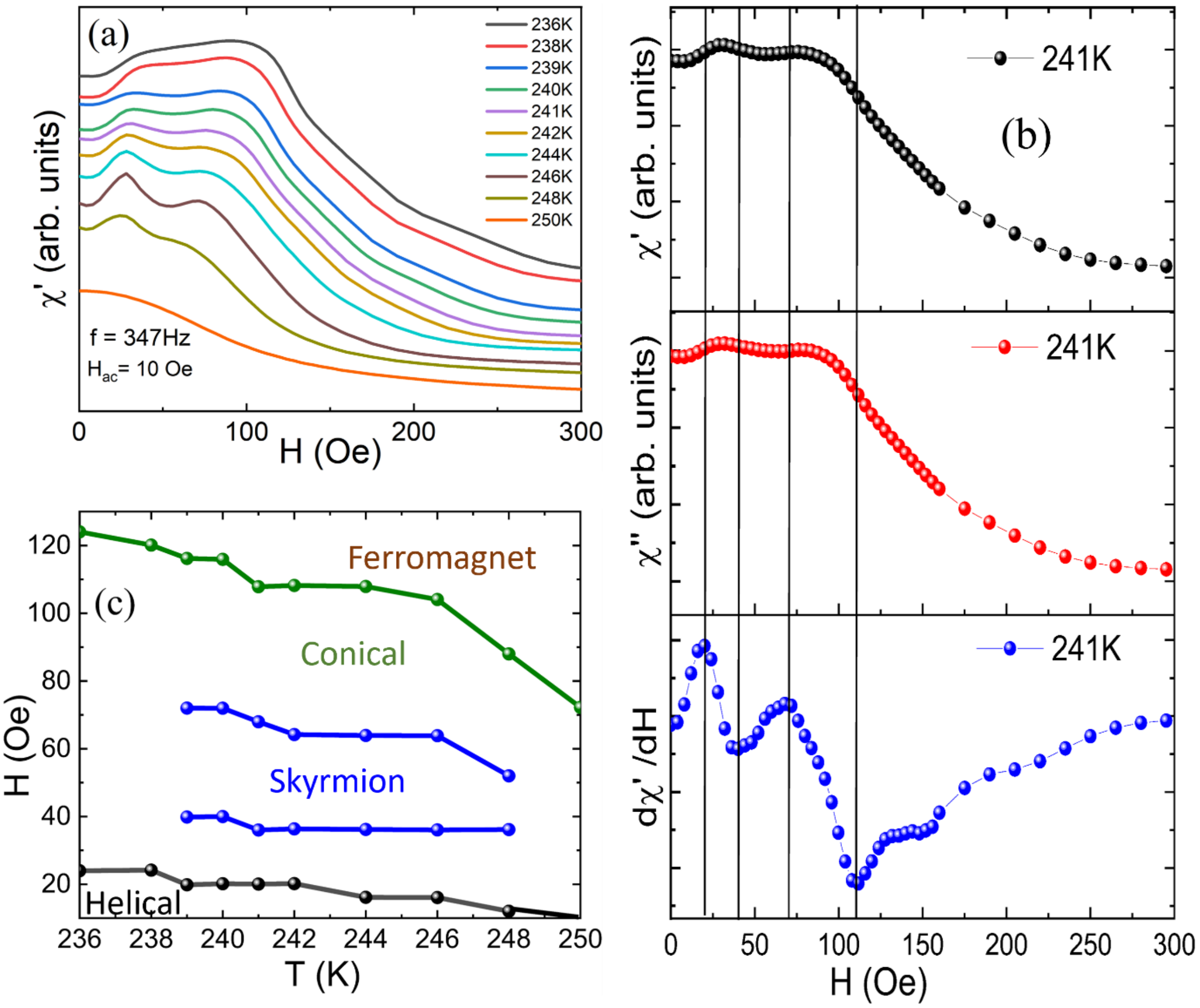

Figure 4:

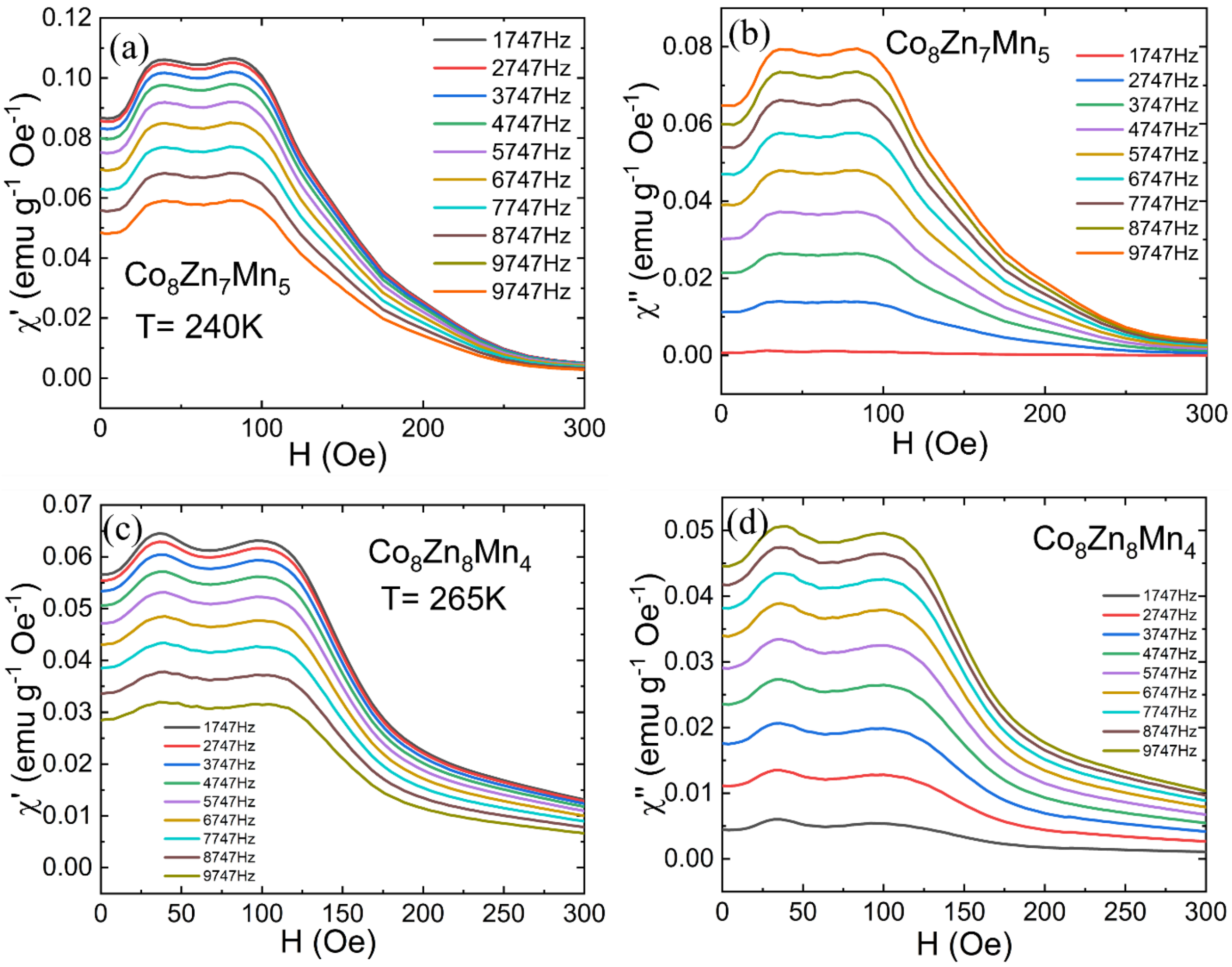

Figure 5

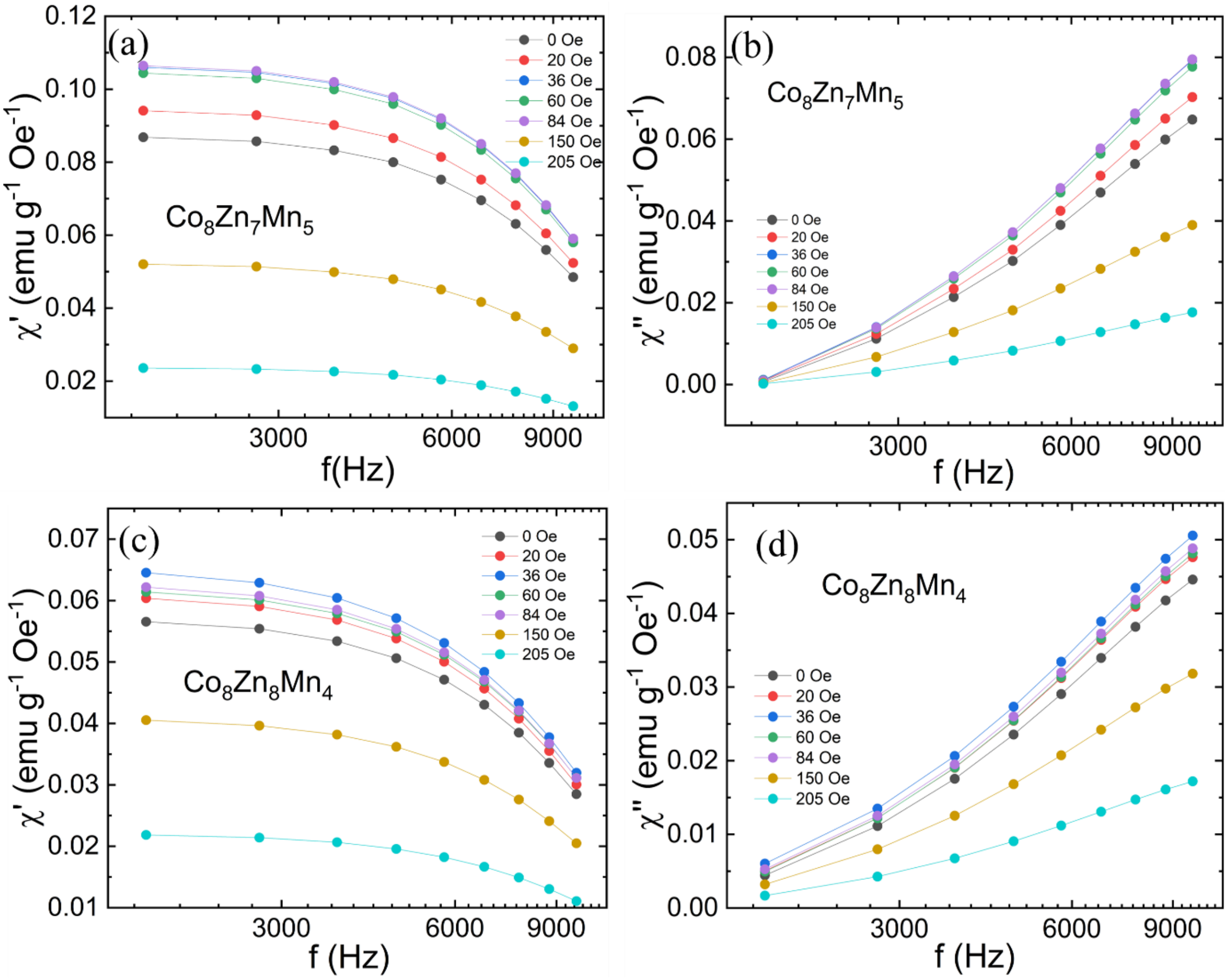

Figure 6

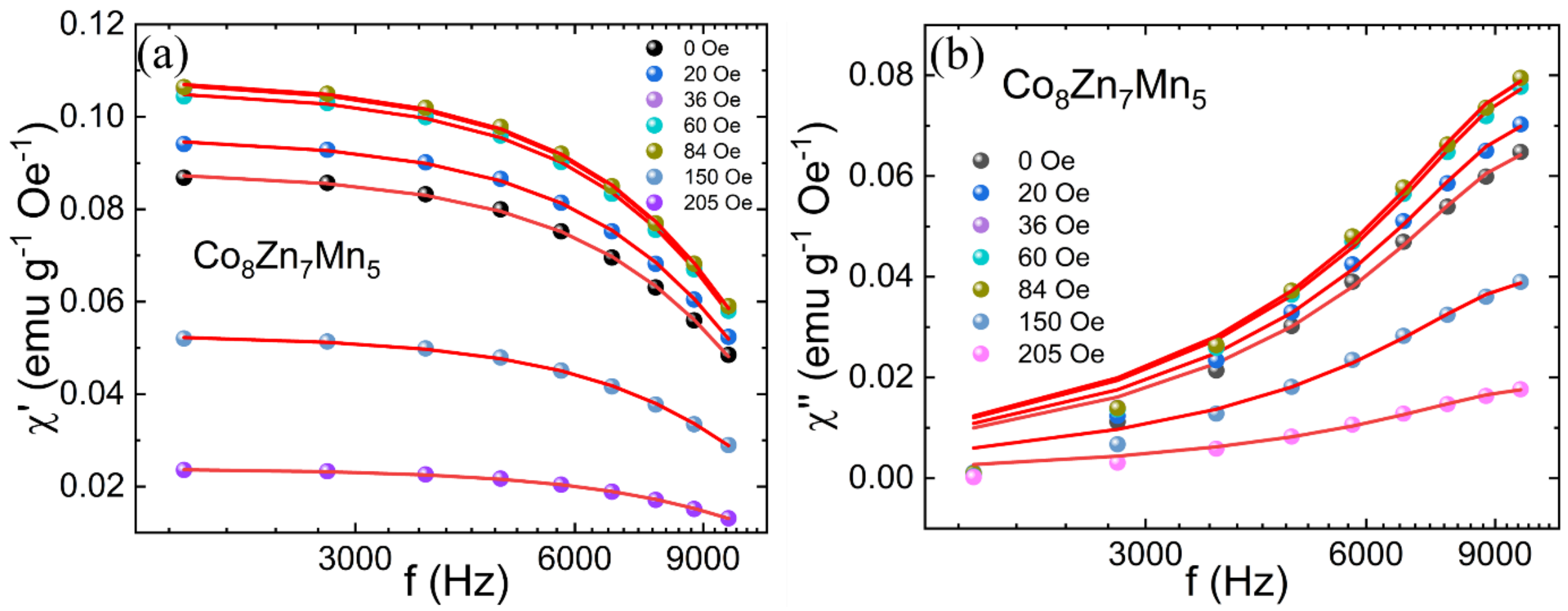

Figure 7

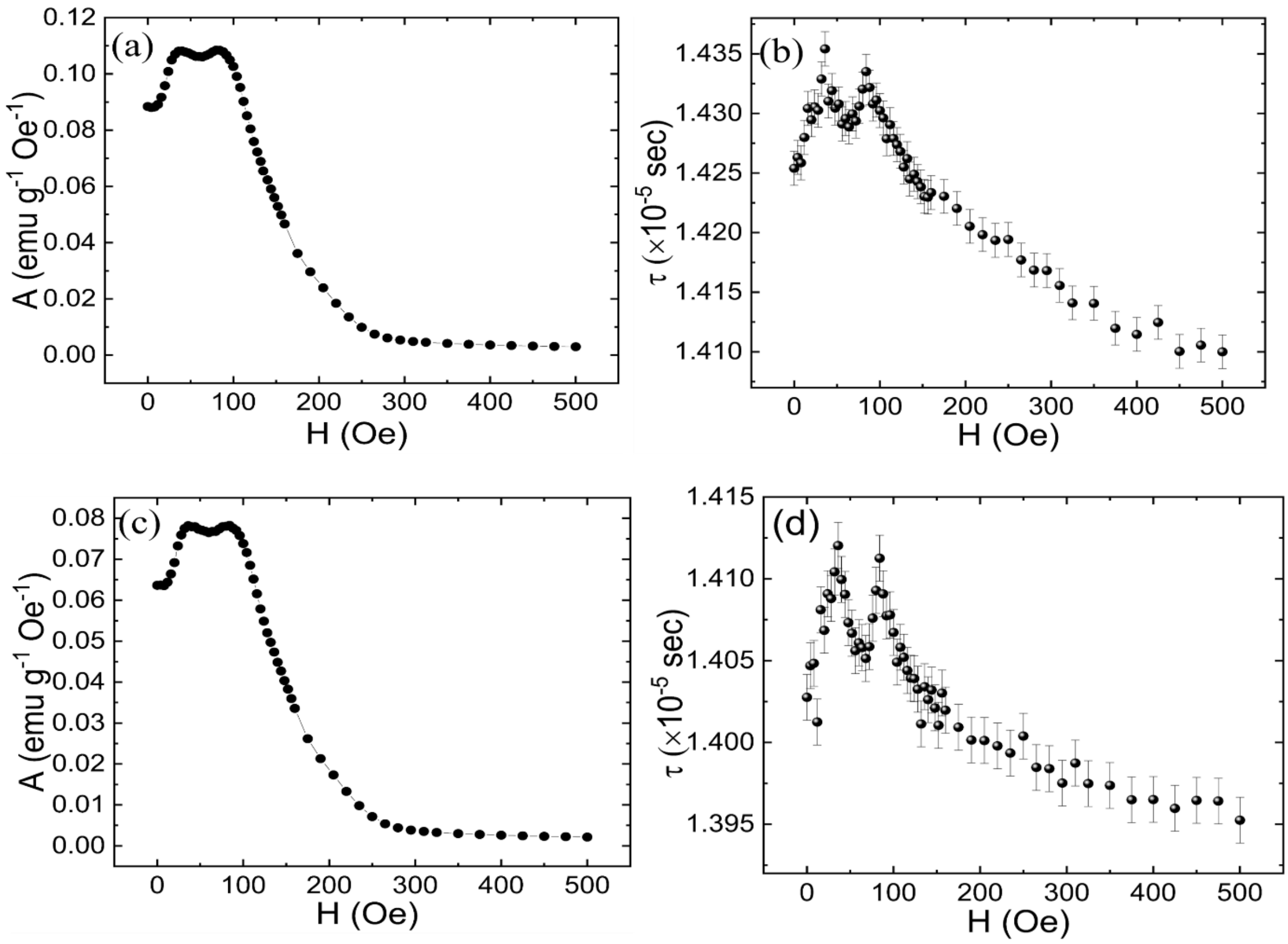

Figure 8

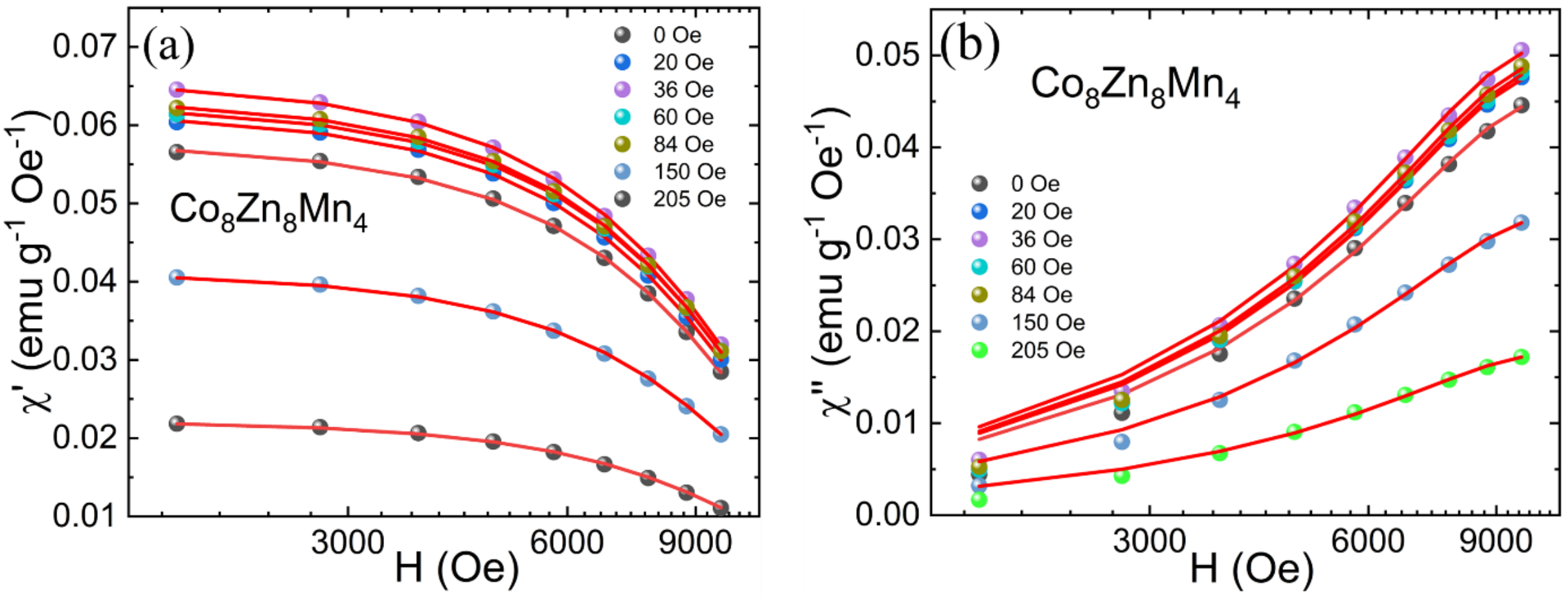

Figure 9

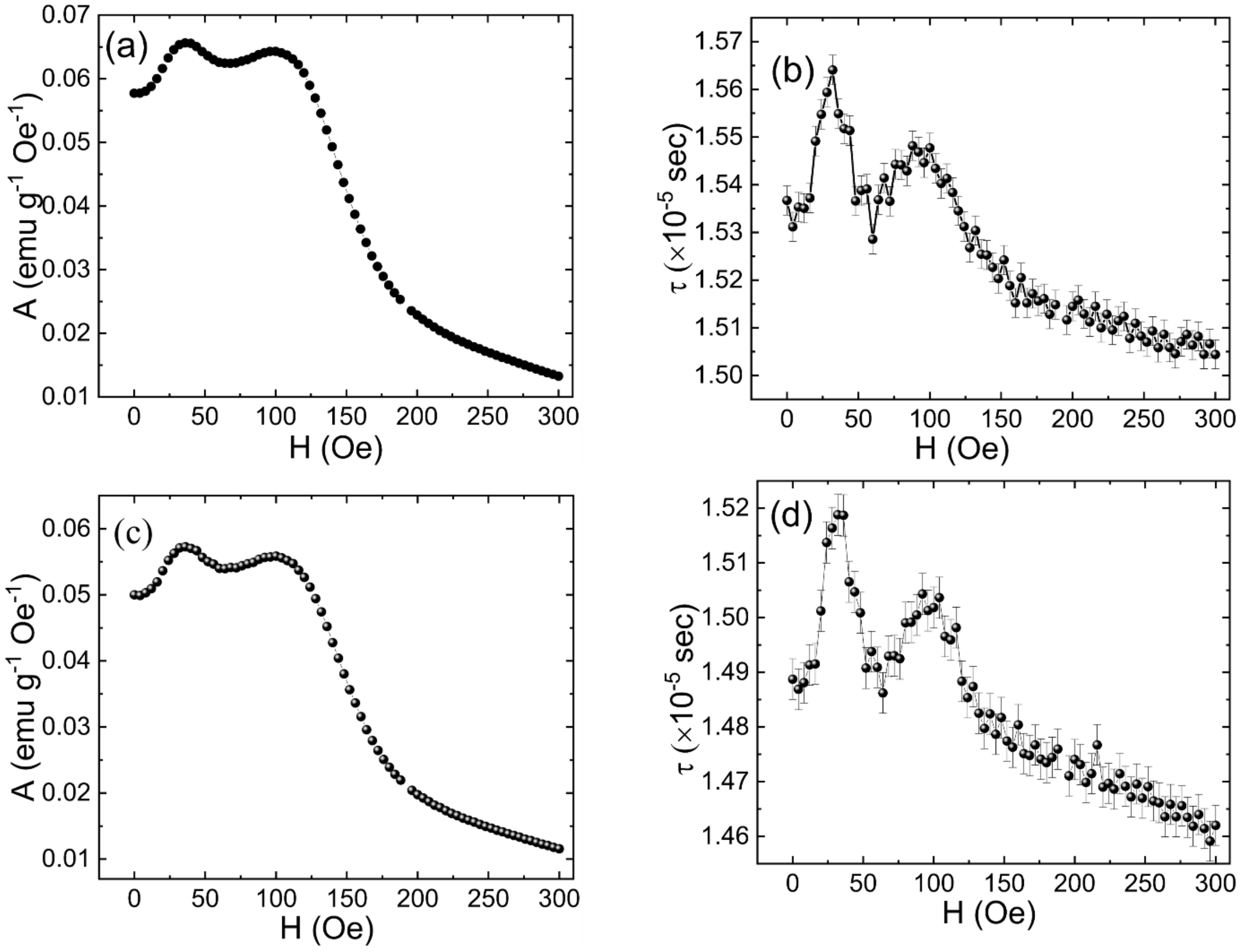